\documentclass[12pt]{article}
\usepackage{graphicx}

\newcommand\pubnumber{Article 18 in eConf C1304143}
\newcommand\pubdate{\today}

\def\nice{ARTEMIS (CNRS/UNS/OCA) UMR 7250\\
Boulevard de l'Observatoire, BP 4229, F-06304 Nice Cedex 4, FRANCE }

\def\Title#1{\begin{center} {\Large #1 } \end{center}}
\def\Author#1{\begin{center}{ \sc #1} \end{center}}
\def\Address#1{\begin{center}{ \it #1} \end{center}}

\newcommand\pubblock{\rightline{\begin{tabular}{l} \pubnumber\\
         \pubdate  \end{tabular}}}
\newenvironment{Abstract}{\begin{quotation}  }{\end{quotation}}
\newenvironment{Presented}{\begin{quotation} \begin{center} 
             PRESENTED AT\end{center}\bigskip 
      \begin{center}\begin{large}}{\end{large}\end{center} \end{quotation}}

\def\beq{\begin{equation}}
\def\eeq#1{\label{#1}\end{equation}}
\def\eeqn{\end{equation}}

\def\beqa{\begin{eqnarray}}
\def\eeqa#1{\label{#1}\end{eqnarray}}
\def\eeqan{\end{eqnarray}}

\let\bar=\overbar

\def\Dslash{\not{\hbox{\kern-4pt $D$}}}
\def\dslash{\not{\hbox{\kern-2pt $\del$}}}

\def\msb{{\bar{\ssstyle M \kern -1pt S}}}

\begin{document}
\begin{titlepage}
\pubblock

\vfill
\Title{Simultaneous detection rates of binary neutron star systems in advanced Virgo/LIGO and GRB detectors}
\vfill
\Author{ Karelle Siellez, \\ Michel Boer and Bruce Gendre}
\Address{\nice}
\vfill
\begin{Abstract}
The coalescence of two compact objects is a key target for the new gravitational wave observatories such as Advanced-Virgo (AdV), Advanced-LIGO (aLIGO) and KAGRA. This phenomenon can lead to the simultaneous detection of electromagnetic waves in the form of short GRBs (sGRBs) and gravitational wave transients. This will potentially allow for the first time access to the fireball and the central engine properties. We present an estimation of the detection rate of such events, seen both by a Swift-like satellite and AdV/ALIGO. This rate is derived only from the observations of sGRBs. We show that this rate, if not very high, predicts a few triggers during the whole life time of Advanced LIGO-Virgo. We discuss how to increase it using some dedicated observational strategies. We apply our results to other missions such as the SVOM French-Chinese satellite project or LOFT. 
\end{Abstract}
\vfill
\begin{Presented}
7th Symposium of Nashville in Huntsville \\
Nashville, United States,  April 14--18 , 2013
\end{Presented}
\vfill
\end{titlepage}
\def\thefootnote{\fnsymbol{footnote}}
\setcounter{footnote}{0}

\section{Introduction}
Thanks to the advanced era of VIRGO [AdV] and LIGO [aLIGO], we will be able, in few years to detect the first gravitational wave [GW]. To understand the physics behind this non-photonic messenger, it will be necessary to observe the progenitor of these events with electromagnetic detectors. Indeed, the merging of two compact objects is one of the most promising and best modeled~\cite{Abbott et al. (2008)} potential source of GWs. These events are also the progenitors of the short Gamma-Ray Bursts [sGRBs]~\cite{Eichler et al.(1989)}, which should produce an extremely intense electromagnetic signal. Short GRBs are rare in the Universe, and the sampled volume is so small that the final detection rate is low~\cite{Coward et al.(2012)} but no strict estimation of the detection rate of an event simultaneously in both windows, based on actual observation, has been done. Most of the results obtained so far were derived from theoretical modeling and population synthesis hypotheses. The aim, here, is to cover this gap, using the most recent observations to estimate that rate. 

We present our selection method in section \ref{Sec_method} and use our final sample to derive the local rate of sGRBs in Section \ref{sec_result}. Then we deduce the rate of simultaneous detections of electromagnetic/gravitational waves events from NS-NS binaries. In section \ref{sec_discu} we discuss our results and their consequences in terms of detectability. We finally conclude in Section \ref{sec_ccl}. In the remainder of this paper, all errors are quoted at $1\sigma$ when not specifically indicated. We use a standard flat cold dark matter ($\Lambda$CDM) model for the Universe, with $\Omega_m = 0.27$, $\Omega_{\Lambda} = 0.73$ and $H_0 = 71$ km s$^{-1}$ Mpc$^{-1}$. We refer the reader to Siellez et al. (2014) \cite{Siellez et al.(2014)} for details. 

\section{Data selection and methods}
\label{Sec_method}

To construct our sample, we used the 679 bursts detected from the \textit{Swift} satellite until June 2012, among which 191 have a known redshift. Unfortunately, the definition of sGRBs is entirely empirical~\cite{Kouveliotou et al.(1993)}, and has no physical ground: sGRBs last less than two seconds {\it in the observer frame} and have harder spectra than long GRBs [lGRBs]. This definition has an obvious limitation: a burst that would be classified as short at a given redshift would be classified as long at a larger redshift, because of time dilation and cosmological effects ~\cite{Kochanek et al. (1993)}. On the contrary, some lGRBs produced by the merger of neutron stars, or black hole - neutron star systems ~\cite{van Putten(2009)} may be considered as short. Thus, we used another discriminative method to separate short and long GRBs using three different filters.  

The first is the rest frame duration. Indeed, as already stated, a short burst risks being confused with a long one in case of high redshift. We thus decided to use the rest frame duration as a first criteria $\tau_{90}$, the 90\% burst duration in the rest frame. We removed from the raw sample all bursts with $\tau_{90} > 2$s.

The second parameter is the spectral selection, because sGRBs are harder than long ones~\cite{Kouveliotou et al.(1993)}. The Band model~\cite{Band et al.(1993)} is a good description of the GRB spectrum. It consists of a broken power law smoothly joined at a typical energy, $E_0$. But the BAT instrument usually detects only one segment of this model. The soft segment is called $\alpha$ (typical value is  1.2), and the hard one is  $\beta$(typical value is 2.3). We have assumed that for a hard burst, the BAT would have detected only $\alpha$ because the peak energy is above the BAT high energy limit. This translates to consider a burst to be hard only if the measured spectral index is lower than 2. We rejected all other events.

The last parameter for selection is the plateau phase which has been discovered by {\em Swift} ~\cite{Tagliaferri et al.(2005)}, and could be due to energy injection~\cite{Zhang et al.(2006)}, a soft tail of a disguised lGRB or produced by a magnetar progenitor~\cite{Metzger et al.(2011)}. As we are interested in the merging of a neutron star binary system (where little energy should be available once the merging is done), we prefer to remove all bursts with a plateau phase, assuming they are related to other kinds of progenitors. Less than the half of the candidates that passed the two previous filters survived to this one. Last, for some rare bursts where the light curve does not allow to determine if a plateau phase is present or not, we relaxed this criteria and validated these events.

\section{Detection rate}
\label{sec_result}

Our final sample consists of 17 events. From this distribution, we estimated the event density as a function of the redshift. Assuming a power-law model, we obtain a best fit power law index of $-1.6 \pm 0.5 $. This is still instrument dependent: {\em Swift} has a field of view of 1.4 steradians, and has been operated for 7.5 years (the duration of the temporal interval where we estimated the total number of events). Taking that into account, we obtain a rate in the local Universe of :  $D = 2.7 \pm 0.9 ~\rm{Gpc}^{-3}~ \rm{y}^{-1}$. We add the supplementary hypothesis that the Universe (within that range) is isotropic and homogeneous, which is not strictly true but a good proxy. We discuss later the impact of a variation of $D$, i.e. adding close-by or distant GRBs, on our results. 

Applying this result, we obtain for AdV (range of 150 Mpc), the isotropic common event rate (within that range): $  R = 0.04 \pm 0.01$  y$^{-1}$. The combination of AdV/aLIGO, which increases the range up to 300 Mpc (T. Regimbau 2013, private communication), leads to $R = 0.3 \pm 0.1$ y$^{-1}$. These numbers are low, and one may wonder if they are accurate. We discuss this point in the next Section, but already note that they are based on a sample detected by {\it Swift}, which is not well suited for detecting sGRBs.

\section{Discussion}
\label{sec_discu}

\subsection{Statistical validation}

\begin{table}[t]
\begin{center}
\begin{tabular}{c|cc|cc}
        &              \multicolumn{4}{c}{Horizon} \\
        &                \multicolumn{2}{c}{AdV} & \multicolumn{2}{c}{AdV/aLIGO} \\
Mission &           R      & N      &             R   &    N        \\
        &y$^{-1}$ & y$^{-1}$ & y$^{-1}$ & y$^{-1}$ \\
\hline
{\it Swift} &  $0.12 \pm 0.04$ & $0.013 \pm 0.004 $& $1.0 \pm 0.3$ & $0.11 \pm 0.04$\\
BATSE   &  $0.4 \pm 0.1$ & $0.10 \pm 0.03$& $ 3.2 \pm 1.1$& $0.8 \pm 0.3$\\
{\it Fermi}-GBM &$0.1 \pm 0.03$ &$0.08 \pm 0.03$ &$0.8 \pm 0.3$ &$0.63 \pm 0.21$\\
LOFT    &$0.07 \pm 0.02$ & $0.02 \pm 0.01$ & $0.6 \pm 0.2$& $0.14 \pm 0.05$ \\
SVOM    & $0.11 \pm 0.04$ & $0.02 \pm 0.01$ & $0.9 \pm 0.3$& $0.14 \pm 0.05$  \\
\hline
\end{tabular}
\caption{Summary of our results for the silver sample (see text): we indicate the detection rate density in volume (D), and the sGRB isotropic event rate (R) and the number of simultaneous EM/GW events per year within the field of view of the instrument (N) for two different ranges: 150 Mpc (AdV detector) and 300 Mpc (AdV/aLIGO combined detectors). 
\label{table_rate_silver}}
\end{center}
\end{table}

\begin{table}[t]
\begin{center}
\begin{tabular}{c|cc|cc}
        &    \multicolumn{4}{c}{Horizon} \\
        &  \multicolumn{2}{c}{AdV} & \multicolumn{2}{c}{AdV/aLIGO} \\
Mission &          R      & N      &             R   &    N        \\
        & y$^{-1}$ & y$^{-1}$ & y$^{-1}$ & y$^{-1}$ \\
\hline
{\it Swift} & $0.04\pm 0.01 $&$ 0.004 \pm 0.001$ &$0.3 \pm 0.1$ & $0.03 \pm 0.01$ \\
BATSE   & $0.12 \pm 0.04$ &$0.03 \pm 0.01$ &$1.0 \pm 0.3$ & $0.24 \pm 0.08$ \\
{\it Fermi}-GBM & $0.03 \pm 0.01$ & $0.023 \pm 0.008$& $0.2 \pm 0.1$ & $0.19 \pm 0.07$\\
LOFT    & $0.02 \pm 0.01$&$0.005 \pm 0.002$ &$0.2 \pm 0.1$&$0.04 \pm 0.01$ \\
SVOM    & $0.03 \pm 0.01$&$0.005 \pm 0.002$ &$0.3 \pm 0.1$&$0.04 \pm 0.01$  \\
\hline
\end{tabular}
\caption{Same as Table 2 for the gold sample (see text). \label{table_rate_gold}}
\end{center}
\end{table}

\begin{figure}
\centering
\includegraphics[width=7cm]{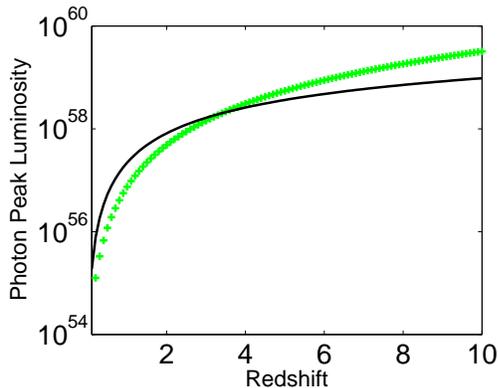}
\caption{Detectability of sGRBs as a function of the redshift. The green crosses represents the \textit{Swift} detection sensibility for a sGRB with all properties (duration, peak flux, Band parameters) set to the median of the observed values. The black solid line represents the peak flux of this template GRB. \label{fig2}}
\end{figure}

In Fig. \ref{fig2}, we compare the peak flux of a template sGRB with the detection threshold of \textit{Swift} for this event. At low redshift this kind of burst can be detected, while at high redshift there are selection effects. We have added a burst at large redshift in our sample and recomputed all rates to find that they remain constant within errors due to the fact that adding a few bursts in this large volume is not significant. At low redshift, we have again inserted a burst in our sample, to find that our rates are multiplied by a factor two. Thus, even if an uncertainty is larger in the final rate, our estimates are still valid. In the following, we will maintain our initial sample, but we will discuss the implications taking into account this uncertainty.

\subsection{Removing the bias}

The first one is the redshift bias. Our sample is based only on sGRBs with a measured redshift, associated to a ``gold sample" of sGRBs that would be detected in the gamma-ray, X-ray, optical bands (i.e. in the electromagnetic spectrum) as well as with gravitational waves. However, only 31.6\% {\em Swift} short bursts have a redshift measurement. We thus define a ``silver sample" of sGRBs that will be detected simultaneously in EM and GW without an associated redshift measurement. We assume that the ratio of sGRB without redshift to the ones with redshift measurement is the same as for canonical sGRBs and  these sGRBs have the same redshift distribution as our ``gold sample". Using these numbers, we find that the rate for the silver sample is $ R=0.12 \pm 0.04$ y$^{-1}$ for a 150 Mpc range (AdV) and $R=1.0 \pm 0.3$ y$^{-1}$ for an range of 300 Mpc (aLIGO/AdV combined).

 There is another bias to correct, due to the sensitivity of {\it Swift} which is not the most suited instrument to detect sGRBs. We have assumed that the discrepancy in sensitivity does not modify the distribution in redshift nor the ratio of sGRBs selected with our method to canonical sGRBs. This last statement means that this ratio of 31.6 \% of sGRBs with a known redshift, is a constant for all missions. The results are given in tables \ref{table_rate_silver} and \ref{table_rate_gold} for the silver and gold samples respectively. The final number of common EM/GW events that can be expected each year is 0.11 when, in the best possible scenario, nowadays, with {\it Fermi}-GBM, we obtain $N = 0.63 \pm 0.21$ sGRB y$^{-1}$ for a 300 Mpc range. The GBM uncertainty on the GRB positions are large. It is thus a key point to be prepared to observe a large portion of the sky with enough sensitivity.

\begin{table}[t]
\begin{center}
\footnotesize
\begin{tabular}{c|c|c}
Work               & Method   &  Estimated GW detection rate \\
                   &          & (y$^{-1}$)\\
\hline
This work,  & Observational constraints &  92 -- 1154 \\
Coward et al.(2012) \cite{Coward et al.(2012)}                             & Observational constraints & 8 -- 1800 \\
Guetta et al. (2006) \cite{Guetta et al. (2006)}                             & Theoretical modeling      & 8 -- 30 \\
Abadie et al. (2010) \cite{Abadie et al. (2010a)}                             & Theoretical modeling      & 2.6 -- 2600 \\
\hline
\end{tabular}
\caption{Predictions of GW detection rates for AdV/aLIGO from this work and comparison with other authors: the first column gives the paper reference, the second the method used by the others, and the last the estimated rate for the AdV/aLIGO network 
\label{table_angle}}
\end{center}
\end{table}
 
\subsection{Comparison with other results}

Previous studies have in general not derived the rate of dual observations, rather the rate of detection of GWs. In order to correct our result of the common EM/GW detection rate into a GW detection rate we have to apply a correction for the beaming angle~\cite{Rhoads(1999)}. We have considered the two extreme measurements of the beaming : $\theta_j = 7^\circ$, for GRB 051221A ~\cite{Soderberg et al.(2006)} and $\theta_j \sim $$25^\circ$ for GRB 050724; using this value for all bursts, we obtain $D_{GW} = 1154 \pm 389$ Gpc$^{-3}$ y$^{-1}$ for the smallest angle and $D_{GW} = 92 \pm 31$ Gpc$^{-3}$ y$^{-1}$  for the largest. Our estimation of $D_{GW}$ is between 92 and 1154, thus, these numbers are in agreements with previous work reported in Table \ref{table_angle}. We thus conclude that our estimates are fair and in good agreement with previous papers.
\section{CONCLUSION}
\label{sec_ccl}

In this presentation, we have presented an estimation of the simultaneous detection of sGRBs and GW events by different electromagnetic and gravitational wave detectors. We have assumed that they both originate from the coalescence of NS-NS binary system. Using the {\it Swift} catalog, we derived a set of 17 sGRBs corrected from instrumental/local effects which has been used to derive the rate density of events expected from present and future GRB missions ({\it Swift}, {\it Fermi}, LOFT, and SVOM) within the range of AdV and the combination of AdV/aLIGO. Even if the common EM/GW detection for which we can expect a distance measurement is low (0.03 per year) for \textit{Swift}, we expect a number close to 1 event for the simultaneous detection of Fermi and AdV/aLIGO combined, which should be observed at all wavelength. Then for the events detected only in gamma-ray and by gravitational waves, these estimations, even if not high, confirm the feasibility of the detection of GWs during the first years of the advanced gravitational wave detector and the common detection with both EM/GW radiations. Planned missions (LOFT and SVOM) will not increase this rate, and in fact {\it Fermi} is more suited for this task due to its larger field of view and higher sensitivity and energy range. The EM follow-up is a way to both confirm a detection and to maximize the science that can be done and the understanding of the sources (neutron stars) as well as the dynamic of the coalescing binary system and its by-product (the sGRB).Preparing a comprehensive set of EM instruments at all wavelengths, is an important objective that should be addressed before Virgo and LIGO start their operational life, i.e. now.


\begin{thebibliography}{99}
\bibitem{Abbott et al. (2008)} Abbott, B., et al., ApJ, 681, 1419 (2008)
\bibitem{Abadie et al. (2010a)} Abadie, J. et al. 2010a, Class. And Quant. Grav., Vol. 17, Issue 17
\bibitem{Band et al.(1993)} Band, D., Matteson, J., Ford, L., et al., 1993, ApJ, 413, 281
\bibitem{Coward et al.(2012)} Coward, D., et al., 2012, MNRAS, 425, 2668
\bibitem{Eichler et al.(1989)} Eichler D. et al., 1989, Nat, 340, 126
\bibitem{Guetta et al. (2006)} Guetta, D. and Piran, T. 2006, A\&A, 453, 823
\bibitem{Kochanek et al. (1993)} Kochanek, C.S. and Piran, T., 1993, ApJ, 417, L17
\bibitem{Kouveliotou et al.(1993)} Kouveliotou, C., Meegan, C.A., Fishman, G.J., et al., 1993, ApJ, 413, L101
\bibitem{Metzger et al.(2011)}	Metzger, B. D., Giannios, D., Thompson, T. A., Bucciantini, N., Quataert, E., 2011, MNRAS 413, 2031
\bibitem{Rhoads(1999)} Rhoads J. E., 1999, ApJ, 525, 737
\bibitem{Siellez et al.(2014)} Siellez K., Boer, M., Gendre, B., 2014, MNRAS, 437, 649
\bibitem{Soderberg et al.(2006)} Soderberg A. M. et al., 2006, ApJ, 650, 261
\bibitem{Tagliaferri et al.(2005)} Tagliaferri, G., et al., 2005, Nat, 436, 985
\bibitem{van Putten(2009)} van Putten, M.H.P.M., MNRAS, 396, L81, 2009
\bibitem{Zhang et al.(2006)} Zhang, B., Fan, Y. Z., Dyks, J., Kobayashi, S., Meszaros, P., Burrows, D.N., Nousek, J.A., Gehrels, N., 2006, ApJ, 642, 354







\end{thebibliography}
\end{document}